\documentclass[10pt,twocolumn,letterpaper]{article}

\usepackage[pagenumbers]{cvpr} 

\usepackage{graphicx}
\usepackage{amsmath}
\usepackage{amssymb}
\usepackage{booktabs}
\usepackage{multirow}
\usepackage{tabularx}
\usepackage{siunitx}
\usepackage{makecell}
\usepackage{float}
\usepackage{placeins}

\usepackage[pagebackref,breaklinks,colorlinks,allcolors=black]{hyperref}

\title{SYNAPSE: Synergizing an Adapter and Finetuning for High-Fidelity EEG Synthesis from a CLIP-Aligned Encoder}

\author{Jeyoung Lee\\
The Catholic University of Korea\\
43 Jibong-ro, Wonmi-gu, Bucheon-si, Gyeonggi-do\\
{\tt\small dlwpdud@catholic.ac.kr}
\and
Hochul Kang\\
The Catholic University of Korea\\
43 Jibong-ro, Wonmi-gu, Bucheon-si, Gyeonggi-do\\
{\tt\small hckang19@catholic.ac.kr}
}

\begin{document}
\maketitle
\begin{abstract}
Recent progress in diffusion-based generative models has enabled high-quality image synthesis conditioned on diverse modalities. Extending such models to brain signals could deepen our understanding of human perception and mental representations. However, electroencephalography (EEG) presents major challenges for image generation due to high noise, low spatial resolution, and strong inter-subject variability. Existing approaches—such as DreamDiffusion, BrainVis, and GWIT—primarily adapt EEG features to pre-trained Stable Diffusion models using complex alignment or classification pipelines, often resulting in large parameter counts and limited interpretability.

We introduce SYNAPSE, a two-stage framework that bridges EEG signal representation learning and high-fidelity image synthesis. 
In Stage1, a CLIP-aligned EEG autoencoder learns a semantically structured latent representation by combining signal reconstruction and cross-modal alignment objectives. In Stage2, the pretrained encoder is frozen and integrated with a lightweight adaptation of Stable Diffusion, enabling efficient conditioning on EEG features with minimal trainable parameters. Our method achieves a semantically coherent latent space and state-of-the-art perceptual fidelity on the CVPR40 dataset, outperforming prior EEG-to-image models in both reconstruction efficiency and image quality. 
Quantitative and qualitative analyses demonstrate that SYNAPSE generalizes effectively across subjects, preserving visual semantics even when class-level agreement is reduced. These results suggest that reconstructing what the brain perceives, rather than what it classifies, is key to faithful EEG-based image generation.
\end{abstract}    
\section{Introduction}
\label{sec:intro}
Recent advances in diffusion-based generative models have enabled high-quality conditional image synthesis across diverse modalities. Stable Diffusion (SD)~\cite{sdv1} established a scalable foundation for multimodal generation, motivating research into image synthesis conditioned on novel signals, including brain activity.

Translating brain signals into visual representations could provide insight into human perception and cognition, with potential applications in mental health and dream interpretation. Most prior work has focused on reconstructing images from functional magnetic resonance imaging (fMRI) signals~\cite{fmri1,fmri2,fmri3}, achieving promising results. However, fMRI acquisition is expensive, slow, and physically demanding, motivating research on Electroencephalography (EEG) as a more accessible alternative.

EEG-based generation presents distinct challenges due to its high noise level, low spatial resolution, and strong inter-subject variability. A core difficulty lies in bridging the modality gap—mapping noisy EEG signals to a structured latent space that captures visual semantics.
Early attempts using Generative Adversarial Networks (GANs)~\cite{gan} produced low-resolution outputs with limited fidelity. DreamDiffusion~\cite{dreamdiffusion} and BrainVis~\cite{brainvis} later investigated EEG-conditioned SD to reconstruct visual experiences. Although these studies opened an important research direction, their designs did not explicitly account for the temporal and structural characteristics of EEG. DreamDiffusion directly projected EEG features into the SD latent space, leading to unstable modality alignment and a large parameter overhead due to MAE pretraining. BrainVis adopted a cascaded diffusion framework that first performs EEG-to-label classification, substitutes each predicted label with a corresponding word token, and encodes it through a frozen CLIP~\cite{clip} text encoder to condition the second-stage diffusion. This indirect conditioning simplifies EEG supervision but restricts the expressiveness of continuous EEG representations, constraining the generation process to discrete label semantics. Moreover, both pipelines employ large parameter counts—MAE pretraining in DreamDiffusion and cascaded diffusion stages in BrainVis—while their reported FID scores remain relatively high.

Existing EEG datasets such as EEGCVPR40~\cite{data1,data2} also provide limited data per subject, particularly for Subject 4, which has been used as a standard benchmark in previous studies~\cite{dreamdiffusion,brainvis,gwit}. Such scarcity and inter-subject variability make single-subject learning prone to overfitting and hinder generalization, further highlighting the need for a multi-subject framework. Recent work such as GWIT~\cite{gwit} leveraged ControlNet~\cite{controlnet} to simplify the training pipeline and achieve high classification accuracy of generated images. However, its comparison on EEGCVPR40 was still limited to a single-subject setting, and the ControlNet-based design reduced interpretability while maintaining relatively high FID scores.

These limitations suggest that previous methods primarily adapted EEG into existing SD frameworks rather than tailoring the generation process to EEG as a structured modality. Additional alignment or classification-based conditioning increases model complexity and obscures how EEG features contribute to image synthesis. This motivates a framework that directly embeds EEG representations into a semantically consistent space shared with CLIP~\cite{clip}, eliminating the need for post-hoc alignment networks.

To address these challenges, we propose SYNAPSE, an EEG-conditioned image generation framework designed for multi-subject learning and efficient training. SYNAPSE incorporates a hybrid autoencoder that extracts compact and interpretable condition vectors with reduced parameters. The encoder is integrated into Stable Diffusion through lightweight selective finetuning, enabling end-to-end training on a single RTX 3090 GPU. Furthermore, by combining IP-Adapter~\cite{ip_adapter} and Classifier-Free Guidance (CFG)~\cite{cfg}, SYNAPSE achieves substantially lower FID scores and improved perceptual fidelity across subjects compared to existing EEG-based models.
\section{Related Work}
\label{sec:related}

This paper is related to two main areas: multimodal image generation and neuroscience-based image generation.

\subsection{Multimodal Image Generation}

The emergence of SD~\cite{sdv1} brought significant advances in multimodal image generation. Its core design leverages Contrastive Language-Image Pre-Training (CLIP)~\cite{clip} for cross-modal representation learning, inspiring numerous Vision-Language Models (VLMs). Subsequent versions, including SD v2~v3, improved fidelity through larger OpenCLIP~\cite{openclip} embeddings, diffusion transformers, and flow-matching~\cite{flowmatching} objectives. SDXL~\cite{sdxl} introduced a two-stage architecture that separates the base and refiner models for better adherence to complex prompts, while SD-Turbo~\cite{sdturbo} applied adversarial diffusion distillation to enable near real-time generation.

Beyond architecture, many approaches focus on efficient finetuning for customized generation. Dreambooth~\cite{dreambooth} enabled subject-specific synthesis, while Low-Rank Adaptation (LoRA)~\cite{lora} introduced parameter-efficient finetuning (PEFT) strategies, later extended by DoRA~\cite{dora} and MoRA~\cite{mora}. However, these methods remain primarily text-conditioned. IP-Adapter~\cite{ip_adapter} addressed image-based conditioning, providing lightweight finetuning that maintains visual consistency and operates efficiently on consumer GPUs (e.g., RTX 3090). Wang et al.~\cite{sf} proposed selective finetuning, adapting specific diffusion components for efficient task-specific learning.

Aligning the generated image with the conditioning signal remains a central challenge. Classifier guidance~\cite{cg} improves conditional adherence but requires an external classifier. CFG~\cite{cfg} achieves similar effects via sampling without extra supervision. Building on these foundations, our framework integrates SD v2.1, CFG, IP-Adapter, and selective finetuning to effectively incorporate EEG-based conditioning.

\begin{figure*}[ht!]
\centering
\includegraphics[width=\textwidth]{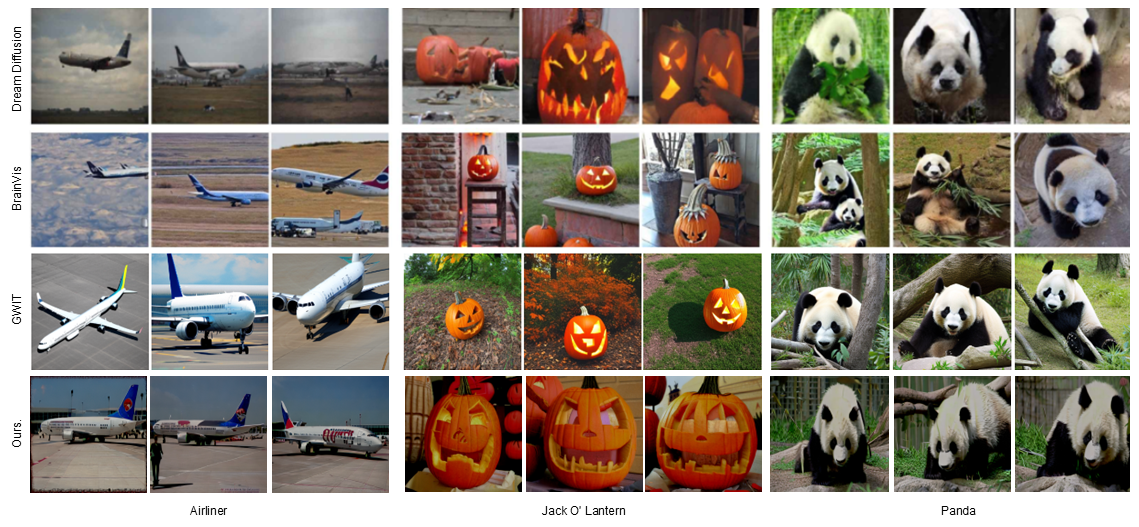}
\caption{
Qualitative comparison of SYNAPSE with state-of-the-art EEG-to-image generation models across representative classes. Each row displays generated samples for 'Airliner', 'Jack O'Lantern', and 'Panda' from a different model (Dream Diffusion~\cite{dreamdiffusion}, BrainVis~\cite{brainvis}, GWIT~\cite{gwit}, and Ours). While prior methods show limitations in structural coherence (e.g., 'Airliner') or semantic fidelity (e.g., 'Panda'), our model (Ours) consistently generates images with higher perceptual quality and closer adherence to the visual features of the original stimuli. This visually supports the state-of-the-art FID score reported in Table~\ref{tab:main_results}.}
\label{fig:Compare_Other}
\end{figure*}

\subsection{Neuroscience-based Image Generation}

Neuroscience-based image generation can be broadly categorized into fMRI-based and EEG-based approaches.
For fMRI, Rakhimberdina et al.~\cite{fmri1} reviewed deep learning techniques for natural image reconstruction, and Huang et al.~\cite{fmri2} surveyed methods for decoding visual information using machine learning. Other studies focused on specific domains, such as face reconstruction~\cite{fmri3}.
These works demonstrate the relative success of fMRI-based image generation.

In contrast, EEG-based models initially received less attention due to significant challenges, including data scarcity and a low signal-to-noise ratio. A foundational dataset for this task, often referred to as EEGCVPR40~\cite{data1,data2}, consists of EEG recordings captured in response to ImageNet stimuli and has served as a primary benchmark. Wang et al.~\cite{eegan} reviewed early GAN-based approaches for this task.Brain2image-GAN~\cite{brain2image_gan} was the first significant attempt to use GANs on the EEGCVPR40 dataset, experimenting with generation via an LSTM-based VAE. Building on this, NEUROVISION~\cite{neurovision} tackled image generation using a Conditional Progressive Growing GAN (PGGAN). Improved-SNGAN~\cite{Improved-SNGAN} enhanced performance by using an encoder combining LSTM and CNN, along with a Spectral Normalization Generative Adversarial Network (SNGAN), while DCLS-GAN~\cite{DCLS-GAN} further improved GAN performance using dual conditioned and lateralization techniques. More recently, EEGStyleGAN-ADA~\cite{adastyle} leveraged the StyleGAN-ADA architecture for training.,While demonstrating feasibility, these methods were limited to low-resolution outputs and suffered from inadequate visual fidelity.

Following the advent of Stable Diffusion, NeuroImagen~\cite{neuroimagen} utilized Stable Diffusion (SD) to control both sampling and conditioning. however, it focused more on caption and image-level control rather than direct matching from EEG signals. DreamDiffusion~\cite{dreamdiffusion} first demonstrated high-resolution synthesis from EEG. However, its framework relied on a single-subject setting and a complex, high-parameter pipeline, including a Masked Autoencoder (MAE)~\cite{mae}, a separate alignment network, and full-model finetuning. BrainVis~\cite{brainvis} aimed to improve this by training an encoder (a hybrid of Ti-MAE~\cite{timae} and an FFT-based model) within cascaded diffusion models. Nevertheless, it still operated in a single-subject setting and did not resolve the issues of complex design, high parameter counts, or the need for a separate alignment network. More recently, GWIT~\cite{gwit} leveraged ControlNet~\cite{controlnet} to achieve real-time performance, lower parameter counts, and a high N-way Top-K Classification Accuracy of Generation (GA) score. However, its comparison on the EEGCVPR40 benchmark was still single-subject. Furthermore, this approach reduced interpretability, as EEG features directly modulate diffusion feature maps without a semantically aligned latent space, and continued to report high FID scores.

To address these limitations, we propose a reconstructive autoencoder that is inherently aligned with the CLIP space, allowing the extraction of detailed and interpretable EEG representations.
Unlike previous frameworks that treat EEG as an auxiliary control signal, our approach explicitly aligns EEG representations with semantic embeddings, enhancing both interpretability and structural fidelity.
Furthermore, we extend our framework from the single-subject setting to a multi-subject environment, providing a foundation for subject-generalized EEG-to-image synthesis.

\begin{figure*}[ht!]
\centering
\includegraphics[width=\textwidth]{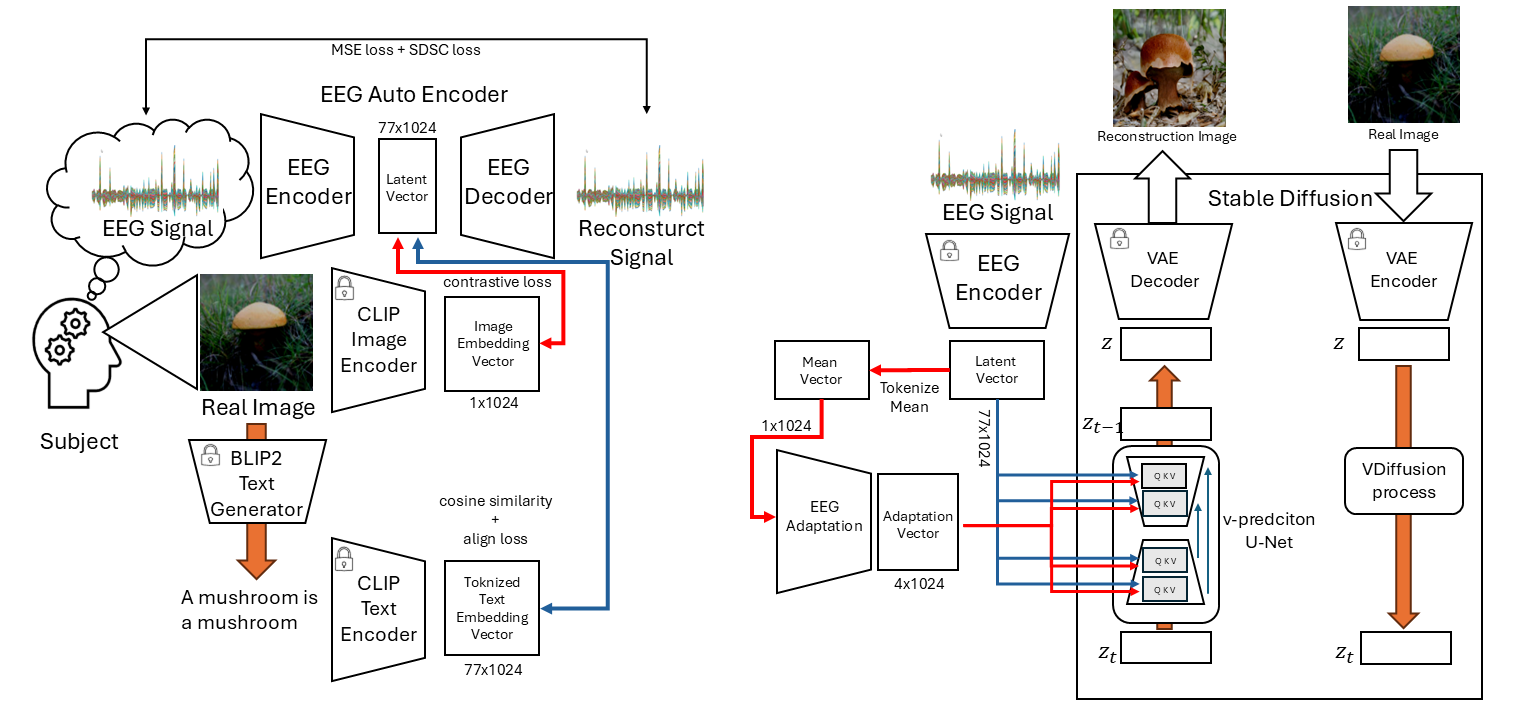}
\caption{
Overview of SYNAPSE, consisting of two stages. 
(Left) Stage 1: Pre-training the hybrid EEG Autoencoder. The encoder $f_{\text{enc}}$ learns to produce a CLIP-aligned latent vector $Z_{\text{latent}}$ by minimizing (1) a reconstruction loss ($L_{\text{recon}}$) for signal fidelity and (2) an alignment loss ($L_{\text{align}} + L_{\text{Contrastive}}$) for semantic consistency with CLIP image–text embeddings. 
(Right) Stage 2: Finetuning Stable Diffusion. The pre-trained encoder $f_{\text{enc}}$ is frozen, while a lightweight EEG Adaptation module $f_{\text{adapt}}$ and selected cross-attention layers of the SD U-Net are finetuned to synthesize images conditioned on both $Z_{\text{latent}}$ and $Z_{\text{adapt}}$.
}
\label{fig:main_framework}
\end{figure*}

\section{Method}
\label{sec:method}

Our framework, SYNAPSE, generates high-fidelity images from multi-subject EEG signals. 
As illustrated in Figure~\ref{fig:main_framework}, SYNAPSE comprises two stages: 
(1) pre-training a CLIP-aligned EEG autoencoder to learn an interpretable latent representation $Z_{\text{latent}}$, and (2) finetuning a pre-trained Stable Diffusion (SD)~\cite{sdv1} model using the frozen encoder and a lightweight EEG adaptation module. 

\begin{figure}[t]
  \centering
  \includegraphics[width=\linewidth]{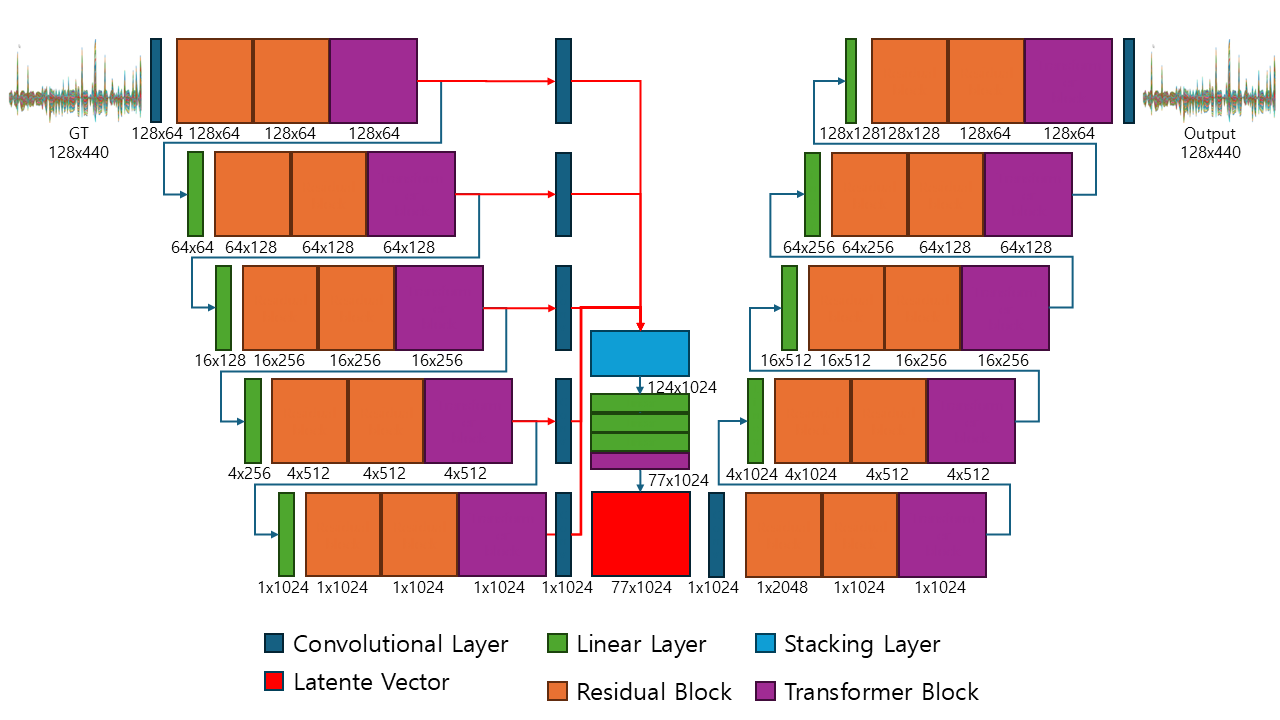}
  \caption{
  Architecture of the hybrid EEG Autoencoder. 
  It employs a U-Net backbone that processes temporal signals through residual blocks and spatial dependencies through transformer blocks to output a CLIP-aligned latent representation.
  }
  \label{fig:autoencoder}
\end{figure}

\subsection{Stage 1: CLIP-Aligned EEG Autoencoder}

The objective of this stage is to train an encoder $f_{\text{enc}}$ that maps a raw EEG signal $E \in \mathbb{R}^{N \times C \times S}$ (batch size $N$, channels $C=128$, time steps $S=440$) into a semantically structured latent space $Z_{\text{latent}} \in \mathbb{R}^{N \times T \times D}$, matching the CLIP text embedding dimension ($T=77$, $D=1024$).

\paragraph{Architecture.}
The encoder $f_{\text{enc}}$ integrates temporal and spatial feature extraction modules within a hierarchical U-Net backbone (Figure~\ref{fig:autoencoder}). Inspired by the analysis of Park et al.~\cite{howdovit}, we apply a similar principle to EEG signals by decomposing the modeling process along temporal and spatial axes rather than spatial scales. Specifically, our hierarchical design first captures per-channel global temporal summaries (spatially local) and then models cross-channel global spatial dependencies (temporally local). This design reinterprets the local-to-global principle in the context of EEG feature composition. Detailed architectures of the Temporal and Spatial Blocks are provided in~\ref{sup:detail_blocks}.

\textbf{Temporal Encoding.}
For each EEG channel, we treat its full temporal sequence ($S$) as a feature vector. A pointwise 1D convolution (Conv1d with kernel\_size = 1) is applied along the temporal axis per channel, performing a learned linear projection from $\mathbb{R}^S$ to $\mathbb{R}^{D_T}$. This operation captures the global temporal dynamics of each channel independently while preserving cross-channel separation.

\textbf{Spatial Encoding.}
The output of the temporal encoder is treated as a set of $C$ tokens, each representing the temporally summarized information of one channel. We apply multi-head self-attention (MHSA) across the channel dimension ($C$) to model global spatial dependencies. Here, each channel $c \in C$ serves as a token, and the attention map models cross-channel relationships. The multi-head mechanism partitions the temporal embedding dimension ($D_T$) into subspaces (e.g., 8 heads), allowing each head to attend to distinct aspects of the temporal representations.

\textbf{Latent Bottleneck.}
The encoded features from both pathways are aggregated through a stacking layer and projected into the target latent space $Z_{\text{latent}} \in \mathbb{R}^{N \times T \times D}$. 
This hierarchical design—per-channel global temporal summary (spatially local) followed by cross-channel global spatial integration (temporally local)—enables the encoder to model EEG signals from fine-grained temporal changes to broader spatial structures, providing a semantically aligned representation compatible with CLIP embeddings.

\paragraph{Loss Functions.}
Training minimizes a multi-objective loss balancing signal reconstruction and semantic alignment.

\textbf{(a) Reconstruction Loss.} 
We combine MSE ($L_{\text{MSE\_recon}}$) and a Signal Dice Similarity Coefficient (SDSC) loss ($L_{\text{SDSC}}$)~\cite{sdsc}.
\begin{equation}
\label{eq:sdsc}
L_{\text{SDSC}} = 1 - \frac{2 \sum_{s} \sigma(E_s \cdot \hat{E}_s) \cdot \min(|E_s|, |\hat{E}_s|)}{\sum_{s} (|E_s| + |\hat{E}_s|)}
\end{equation}
where $\sigma(\cdot)$ is the sigmoid function. SDSC measures the overlap of signal peaks between $E$ and $\hat{E}$, encouraging structural consistency.
\begin{equation}
L_{\text{recon}} = \|E - \hat{E}\|_2^2 + \lambda_{\text{sdsc}} L_{\text{SDSC}}(E, \hat{E})
\end{equation}

\textbf{(b) CLIP Alignment Loss.} 
To align $Z_{\text{latent}}$ with the CLIP text embedding $Z_{\text{text}}$, we minimize a combination of MSE ($L_{\text{MSE\_align}}$) and cosine similarity loss ($L_{\text{CosSim}}$):
\begin{equation}
\label{eq:text_align}
\begin{split}
L_{\text{text\_align}} = & \;\lambda_{\text{mse}} \|Z_{\text{latent}} - Z_{\text{text}}\|_2^2 \\
                      & + \lambda_{\text{cos}} (1 - \text{sim}(Z_{\text{latent}}, Z_{\text{text}}))
\end{split}
\end{equation}
where $\text{sim}(\cdot)$ denotes cosine similarity.

\textbf{(c) Contrastive Loss.} 
To ensure cross-modal semantic consistency, we use an InfoNCE-style loss between the mean-pooled latent $\bar{Z}_{\text{latent}}$ and CLIP image embeddings $Z_{\text{image}}$:
\begin{equation}
\label{eq:contrastive}
L_{\text{cont}} = -\frac{1}{N} \sum_{n} \log \frac{\exp(\text{sim}(\bar{Z}_{n}, Z_{\text{image},n}) / \tau)}{\sum_j \exp(\text{sim}(\bar{Z}_{n}, Z_{\text{image},j}) / \tau)}
\end{equation}

\textbf{(d) Overall Loss.}
The final loss for Stage 1 combines these components:
\begin{equation}
\label{eq:stage1_loss}
\begin{split}
L_{\text{Stage1}} = & \;\lambda_{\text{recon}} L_{\text{recon}} + \lambda_{\text{align}} L_{\text{text\_align}} \\
                  & + \lambda_{\text{con}} L_{\text{cont}}
\end{split}
\end{equation}
where $\lambda_{(\cdot)}$ are hyperparameter weights for each loss term.

The CLIP-aligned latent $Z_{\text{latent}}$ obtained from Stage 1 serves as the semantic condition for image generation in Stage 2.

\subsection{Stage 2: EEG-Conditioned Stable Diffusion}

In this stage, we finetune a pre-trained SD v2.1 model using the frozen encoder $f_{\text{enc}}$ from Stage 1 and a lightweight EEG adaptation module $f_{\text{adapt}}$.

\paragraph{Finetuning Strategy.}
We condition the SD U-Net on EEG representations while maintaining computational efficiency (Figure~\ref{fig:main_framework}, right).  
The model is optimized with the v-prediction loss $L_v$~\cite{progressive-distillation}:
\begin{equation}
\label{eq:v_loss}
\begin{split}
L_v = & \;\mathbb{E}_{x_0,\epsilon,t,c} [w(t)\|v_t - v_{\theta}(x_t, t, c)\|^2], \\
    & \quad \text{where } w(t)=\text{SNR}(t)^{-\gamma}
\end{split}
\end{equation}
where $\gamma=0.5$ ensures stability during training.

\paragraph{EEG Adaptation Module.}
Following IP-Adapter~\cite{ip_adapter}, $f_{\text{adapt}}$ projects the mean-pooled latent $\bar{Z}_{\text{latent}} \in \mathbb{R}^{N \times 1 \times D}$ into an adaptation vector $Z_{\text{adapt}} \in \mathbb{R}^{N \times 4 \times D}$ via a lightweight Linear–LayerNorm–Linear design.

\paragraph{Selective Finetuning.}
The U-Net’s cross-attention blocks are conditioned on $c = \text{Concat}(Z_{\text{latent}}, Z_{\text{adapt}})$.  
To prevent catastrophic forgetting and ensure efficiency, all SD weights and $f_{\text{enc}}$ parameters are frozen.  
Only the EEG adaptation module ($f_{\text{adapt}}$) and the key/value projections in the cross-attention layers are updated.

\paragraph{Classifier-Free Guidance (CFG).}
To enhance conditional adherence during inference, we apply Classifier-Free Guidance (CFG)~\cite{cfg}, 
which scales the conditional prediction relative to the unconditional one through a guidance factor $s$:
\begin{equation}
\begin{split}
\tilde{v}_{\theta}(x_t,t,c) = & \; v_{\theta}(x_t,t,\emptyset) + s \cdot (v_{\theta}(x_t,t,c) \\
& - v_{\theta}(x_t,t,\emptyset))
\end{split}
\end{equation}
where $\emptyset$ denotes the unconditional path obtained by randomly dropping EEG conditions.

This two-stage design, combined with selective finetuning and mixed-precision training, enables end-to-end optimization on a single consumer GPU (e.g., RTX 3090).

\section{Experimental Results}

All experiments are conducted on the \textbf{CVPR40} dataset~\cite{data1, data2}, a standard benchmark for EEG-based visual recognition. This dataset contains 128-channel EEG recordings from six participants viewing 2,000 ImageNet-derived images across 40 object categories. Following the established preprocessing protocol, each 0.5-second segment is filtered between 5–95~Hz, the initial 20~ms are discarded, and all signals are uniformly truncated to 440 samples. 
For evaluation, we use a 9:1 split for single-subject and an 8:1:1 split for multi-subject training and validation.

\subsection{Stage 1: Autoencoder Pre-training}

We first evaluate the autoencoder’s capability to reconstruct EEG signals while maintaining semantic consistency.  Our hybrid autoencoder is compared against the MAE-based architecture from DreamDiffusion~\cite{dreamdiffusion}, which also includes a decoder for signal reconstruction.

\paragraph{Reconstruction Efficiency.}
Previous work~\cite{dreamdiffusion} employed a separate, high-capacity alignment network on top of the EEG encoder, which often disrupts the original latent representation. To establish a fair comparison of the encoder backbones, we first assess reconstruction performance when compressing to the same CLIP-aligned latent dimension ($77\times1024$), prior to applying semantic alignment losses.

 \begin{table}[htbp] 
  \centering 
  \caption{ Comparison of EEG autoencoder pre-training methods. 
  Our model demonstrates superior reconstruction fidelity with significantly fewer parameters.
  Lower MSE (↓) and higher SDSC (↑) indicate better performance.
  Detailed results are provided in~\ref{sup:Autoencoder_full}.} 
  \label{tab:main_ae_results} 
  \small 
  \begin{tabular}{lccc} 
    \toprule
    Model                          & {Params (M) $\downarrow$} & {MSE $\downarrow$}        & {SDSC $\uparrow$}       \\
    \midrule
    MAE ~\cite{dreamdiffusion}     & 166.45         & 0.2403         & 0.7590         \\
    \midrule
    Ours (w/o align)               & \textbf{44.63}          & 0.0810         & 0.7893         \\
    \bfseries Ours (CLIP-Aligned)  & \textbf{44.63}          & 0.0881        & 0.7860         \\
    \bottomrule
  \end{tabular}
\end{table}

As shown in Table~\ref{tab:main_ae_results}, our model achieves markedly lower MSE and higher SDSC compared to the MAE baseline, while using approximately 3.7$\times$ fewer parameters (44.6M vs. 166.4M). 
Even after applying CLIP alignment losses, reconstruction quality remains stable, indicating that the alignment process preserves signal fidelity while enhancing semantic structure in the latent space.

\paragraph{Semantic Consistency.}
We next evaluate the semantic properties of the learned latent space $Z_{\text{latent}}$ through Top-$k$ label retrieval and t-SNE visualization. 

\begin{table}[h!]
  \centering
  \small
  \caption{Label Top-$k$ retrieval accuracy of the EEG encoder.
  Multi-subject training substantially improves semantic alignment.}
  \label{tab:main_label_results}
  
  \begin{tabularx}{\linewidth}{@{} l *{3}{>{\centering\arraybackslash}X} @{}} 
    \toprule
    Subject & Top-1 $\uparrow$ & Top-3 $\uparrow$ & Top-5 $\uparrow$ \\
    \midrule
    4         & 0.4330 & 0.6515 & 0.8112\\
    \midrule
    General   & \textbf{0.6022} & \textbf{0.8742} & \textbf{0.9518}\\
    \bottomrule
  \end{tabularx}
\end{table}

Table~\ref{tab:main_label_results} reports the label Top-$k$ retrieval performance. For each EEG latent vector $\bar{Z}_{\text{latent}}$, we compute cosine similarity with all CLIP image embeddings $Z_{\text{image}}$ in the test set and verify whether any of the top-$K$ retrieved images share the same class label. 
The multi-subject model achieves a Top-5 accuracy of 95.2\%, demonstrating robust semantic alignment and improved generalization across subjects.

\begin{figure}[htbp]
  \centering
  \includegraphics[width=\linewidth]{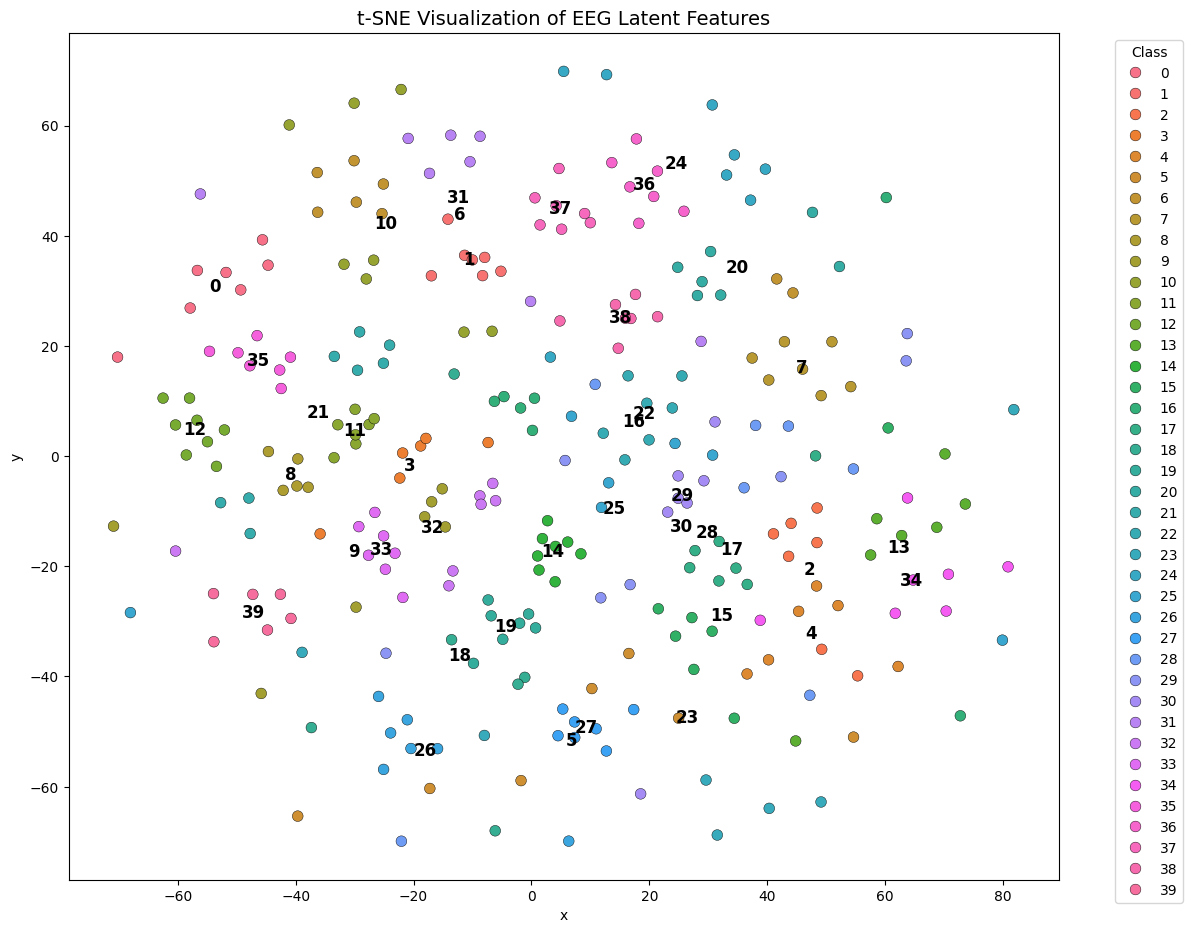}
  \caption{
  Subject 4 T-SNE Example
  }
  \label{fig:tsne_sub4}
\end{figure}

\begin{figure}[htbp]
  \centering
  \includegraphics[width=\linewidth]{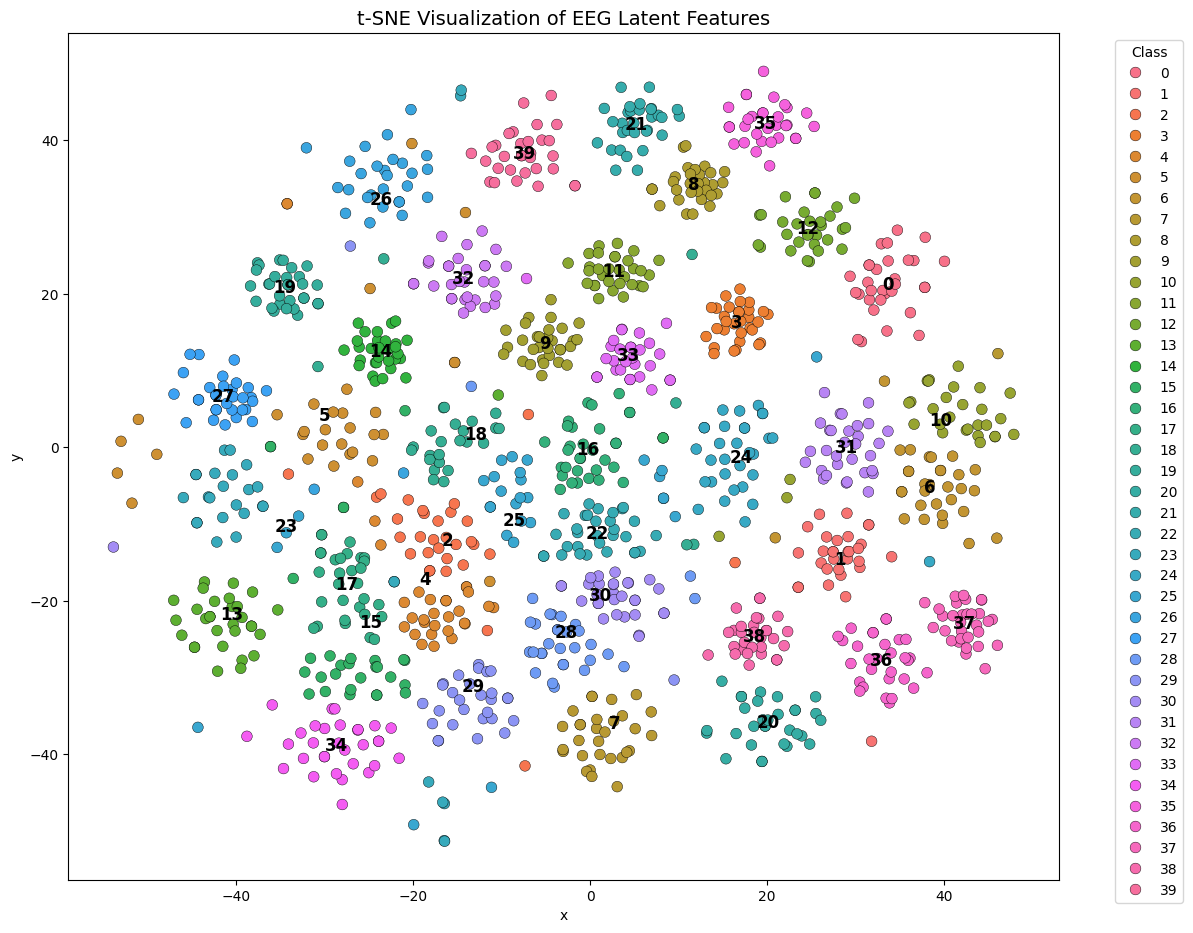}
  \caption{
  Multi Subject T-SNE Example
  }
  \label{fig:tsne_multi}
\end{figure}

\begin{figure*}[t!]
  \centering
  \includegraphics[width=\textwidth]{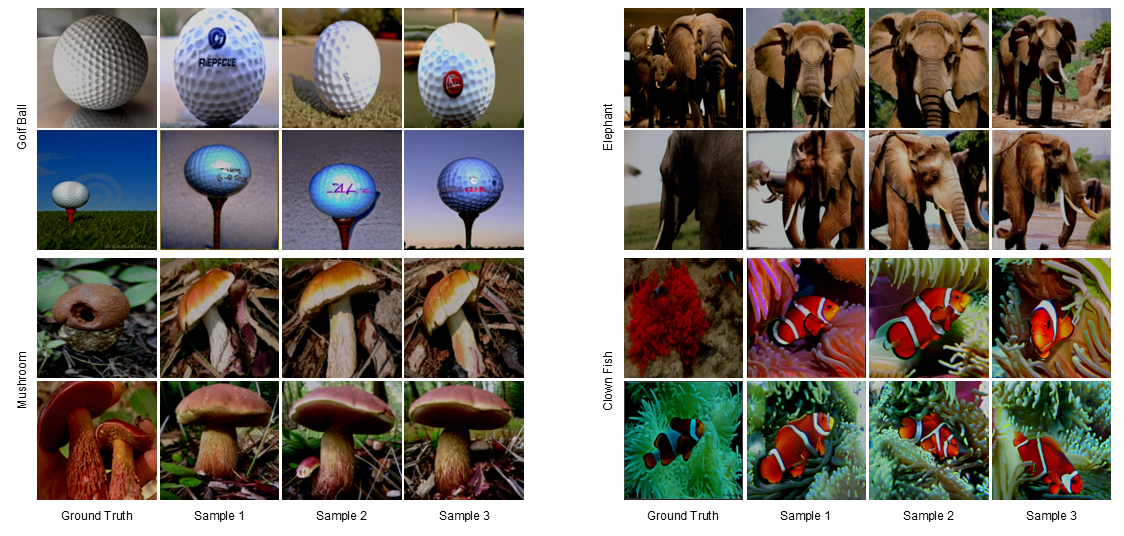}
  \caption{
  Representative examples demonstrating semantic coherence.
  }
  \label{fig:semantic_exam}
\end{figure*}

The t-SNE visualizations in Figures~\ref{fig:tsne_sub4} and~\ref{fig:tsne_multi} support these findings.
While the single-subject model produces scattered and weakly defined clusters, the multi-subject model exhibits clear class separation, confirming that semantic alignment benefits from increased data diversity (approximately 13,000 samples versus 2,800 for Subject~4).

\subsection{Stage2}

We now evaluate SYNAPSE in the full image generation setting, comparing against state-of-the-art EEG-to-image models.

\paragraph{CFG Scaling.}
During inference, we apply CFG with a scaling factor of 7.5. We empirically tested values of 3, 5, 7, and 9, and observed that the best balance between fidelity and diversity occurs around 7–9. Following convention in diffusion-based synthesis, we adopt 7.5 as the default setting. Additional details are provided in~\ref{sup:cfg_detail}.

\paragraph{Quantitative Comparison.}
Table~\ref{tab:main_results} summarizes quantitative performance.
Our final model (\textit{Ours-IP+CFG, Multi}) achieves the best FID, representing nearly a twofold improvement over the previous best reported by GWIT~\cite{gwit}. Inception Score (IS) remains comparable, indicating that improved fidelity does not come at the expense of diversity.

\begin{table}[t] 
  \centering 
  \caption{Comparison of EEG-to-image synthesis models.
  Lower FID (↓) and higher GA, IS (↑) indicate better performance.} 
  \label{tab:main_results} 
  \small 
  \resizebox{\linewidth}{!}{
  \begin{tabular}{llccc} 
    \toprule 
    Model                 & Subject       & FID ↓          & GA ↑           & IS ↑           \\
    \midrule 
    DreamDiffusion ~\cite{dreamdiffusion} & 4 & -              & 0.46          & -  \\
    BrainVis ~\cite{brainvis} & 4             & 121.02         & 0.49          & 31.58        \\
    GWIT ~\cite{gwit}         & 4             & 80.47          & \textbf{0.91}   & \textbf{33.32}  \\
    \midrule
    Our-base           & 4             & 133.36       & 0.21         & 20.00        \\
    Our-IP             & 4             & 148.08       & 0.24         & 19.65        \\
    Our-IP+CFG         & 4             & 116.57       & 0.26         & 23.30        \\
    \midrule 
    Brain2Image-GAN~\cite{brain2image_gan} & Average  & -      & -             & 5.07\\ 
    NeruoVision~\cite{neurovision} & Average          & -      & -             & 5.15\\ 
    Improved-SNGAN~\cite{Improved-SNGAN}& Average     & -      & -             & 5.53\\ 
    DCLS-GAN~\cite{DCLS-GAN} & Average                & -      & -             & 5.64\\ 
    NeuroImagen~\cite{neuroimagen} & Average          & -      & 0.85          & 33.50\\ 
    EEGStyleGAN-ADA~\cite{adastyle} & Average         & 174.13 & -             & 10.82\\ 
    BrainVis ~\cite{brainvis} & Average               & 126.66 & 0.45          & 31.00        \\
    GWIT ~\cite{gwit}         & Average               & 80.47  & \textbf{0.91} & \textbf{33.87}  \\
    \midrule 
    
    Our-base           & Multi         & 72.89        & 0.33         & 27.18         \\
    Our-IP             & Multi         & 72.83        & 0.35         & 30.14        \\
    \textbf{Our-IP+CFG}& \textbf{Multi}&\textbf{46.91}& 0.39         & 31.49        \\
    \bottomrule 
  \end{tabular}}
\end{table}

\paragraph{Discussion on FID–GA Trade-off.}
Our GA score (0.39) is notably lower than that of GWIT (0.91), which is expected given the different design objectives. 
Classification-based models such as GWIT optimize for label accuracy, often producing canonical class exemplars rather than instance-specific reconstructions.
In contrast, SYNAPSE prioritizes perceptual realism by learning continuous visual features directly from EEG signals. 
This design improves fidelity (lower FID) but may reduce class-level agreement (GA).
Given that the goal of EEG-to-image generation is to reconstruct perceptual content rather than predict discrete categories, we argue that FID provides a more meaningful evaluation metric.
The strong semantic alignment observed in Stage~1 further supports this claim.

\paragraph{Qualitative Analysis.}
Figure~\ref{fig:semantic_exam} presents representative qualitative results.
The model produces visually coherent samples that capture both contextual and semantic variations of the original stimuli.
For instance, in the Golf Ball category, SYNAPSE not only reproduces the ball itself but also reliably renders the supporting tee beneath the ball, indicating sensitivity to object–support relations.
In Mushroom, the generations primarily reflect color-specific cues (e.g., cap hue and tone) rather than mere silhouette, adapting the overall palette to the observed chromatic context.
Similarly, for Clown Fish, the model distinguishes characteristic color schemes—most notably red versus blue surroundings—while preserving the fish’s stripe patterns.
Lastly, Elephant samples exhibit clear differentiation in viewpoint—frontal or side-facing—and preserve trunk and tusk geometry.
These examples suggest that SYNAPSE captures fine-grained visual attributes, especially color and local context, beyond categorical identity.

\begin{figure}[htbp]
  \centering
  \includegraphics[width=\linewidth]{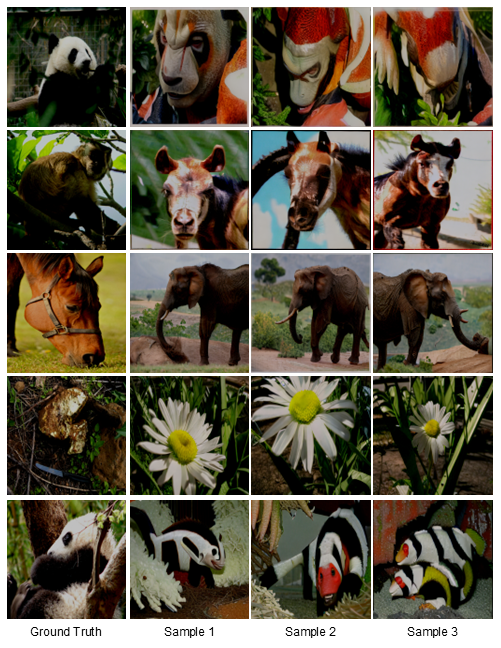}
  \caption{
  Bad case example
  }
  \label{fig:bad_exam}
\end{figure}

\begin{figure}[htbp]
  \centering
  \includegraphics[width=\linewidth]{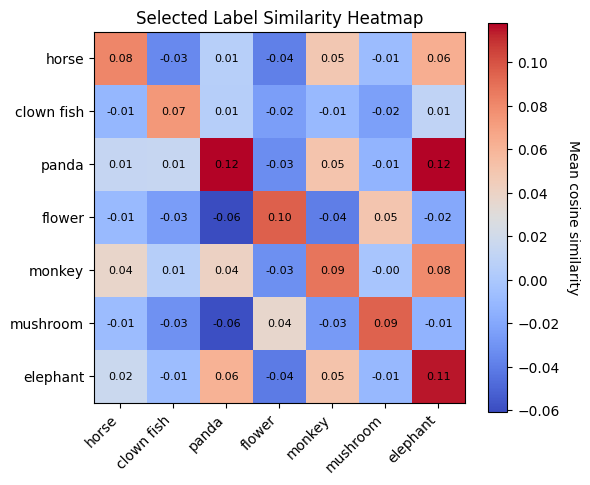}
  \caption{
  Cosine similarity map among selected EEG-derived latent embeddings.
  Warmer colors indicate higher similarity. 
  The plot highlights structured cross-class relations—particularly between semantically related categories 
  }
  \label{fig:cos_sel}
\end{figure}

\paragraph{Failure and Cross-Class Similarity Analysis.}
Figure~\ref{fig:bad_exam} illustrates representative failure cases, which typically arise when the Stage~1 encoder produces ambiguous semantic embeddings, resulting in lower GA. Despite incorrect categorical predictions (e.g., “Horse”→“Elephant”, “Panda”→“Masked Creature”), the generated images often retain coarse structural or textural traits—such as animal morphology, limb proportion, or characteristic color regions—indicating that SYNAPSE maintains perceptual alignment even when categorical accuracy declines.

\begin{figure}[htbp]
  \centering
  \includegraphics[width=\linewidth]{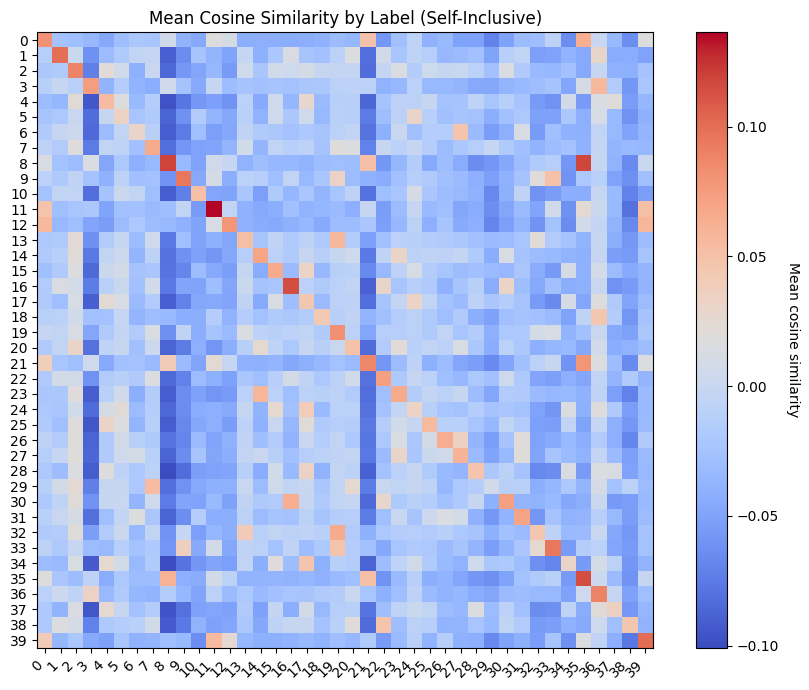}
  \caption{Representative good-case examples. SYNAPSE generates perceptually coherent and semantically aligned images.}
  \label{fig:cos_all}
\end{figure}

To further analyze this phenomenon, Figure~\ref{fig:cos_all} and Figure~\ref{fig:cos_sel} present the cosine similarity among EEG-derived latent embeddings. Although categorical separability remains limited, intra-class coherence is consistently stronger than inter-class similarity. Semantically related categories, such as panda–elephant and horse–monkey, exhibit elevated cross-class similarity, mirroring the visually plausible yet label-ambiguous generations seen in Figure~\ref{fig:bad_exam}. Interestingly, certain misalignments, such as the “masked panda” resembling a clown fish, correspond to regions of high cross-class similarity in the latent space, suggesting that these perceptual confusions emerge from structured overlap rather than random noise. Rather than implying that the subject “perceived an elephant,” these cases suggest that the model successfully extracted low- and mid-level visual cues (e.g., four-legged shape, viewpoint, color tone) from the “horse” EEG but failed to resolve the high-level semantic distinction, leading to a collapse toward a visually similar class. Overall, the results indicate that SYNAPSE preserves a semantically organized latent geometry aligned with visual structure, even when categorical boundaries blur. Additional qualitative comparisons and extended case studies, including good-case examples, are provided in the Supplementary Material (see Section~\ref{sup:more_examples}).

\section{Conclusion}

We presented SYNAPSE, a two-stage framework that unifies EEG signal encoding with diffusion-based image synthesis. By integrating a CLIP-aligned EEG autoencoder with a lightweight adaptation of Stable Diffusion, our method learns a semantically structured latent space and achieves state-of-the-art perceptual fidelity on the CVPR40 benchmark. Extensive experiments demonstrate that SYNAPSE generalizes effectively across subjects while preserving meaningful visual correspondence, outperforming prior methods in both reconstruction efficiency and visual realism.

Our analyses further reveal that the model captures low- and mid-level visual cues—such as shape, viewpoint, and color—directly from EEG signals, even when high-level categorical accuracy declines.
This behavior indicates that apparent label mismatches arise not from random noise but from structured feature overlap in the latent space, suggesting that SYNAPSE reconstructs what the subject perceives rather than what the label describes. Future work will extend this framework toward finer temporal modeling, adaptive subject calibration, and multimodal generalization, moving closer to real-time brain-to-vision translation.

\FloatBarrier  

\clearpage
{
    \small
    \bibliographystyle{ieeenat_fullname}
    \bibliography{main}
}

\clearpage
\setcounter{page}{1}
\maketitlesupplementary

\vspace{1em}
\section{EEG Autoencoder Block Details}
\label{sup:detail_blocks}

\subsection{Temporal Block}
\begin{figure}[htbp]
\centering
\includegraphics[width=0.75\linewidth]{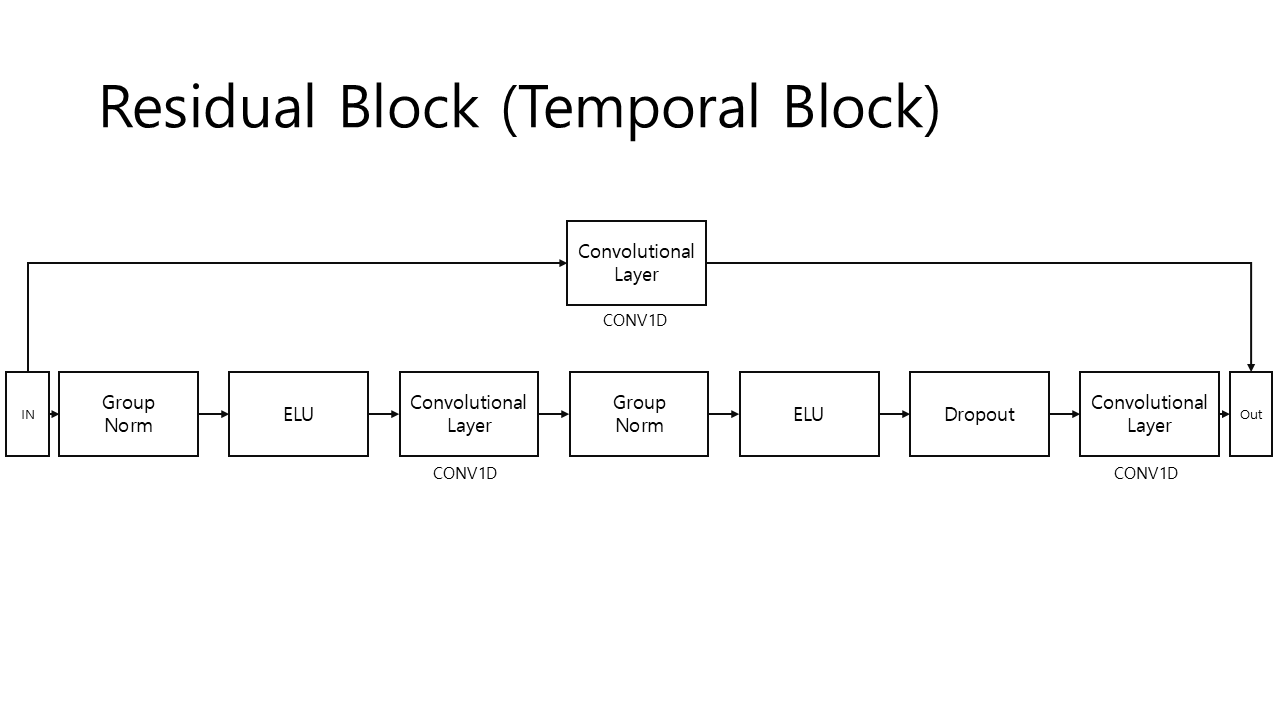}
\caption{Temporal Block structure used in $f_{\text{enc}}$.}
\label{fig:temporal_block_sup}
\end{figure}

Figure~\ref{fig:temporal_block_sup} shows the structure of the Temporal Block, 
consisting of a 1D convolution–BatchNorm–ReLU sequence with a residual connection to preserve local spatial continuity.

\subsection{Spatial Block}
\begin{figure}[htbp]
\centering
\includegraphics[width=0.75\linewidth]{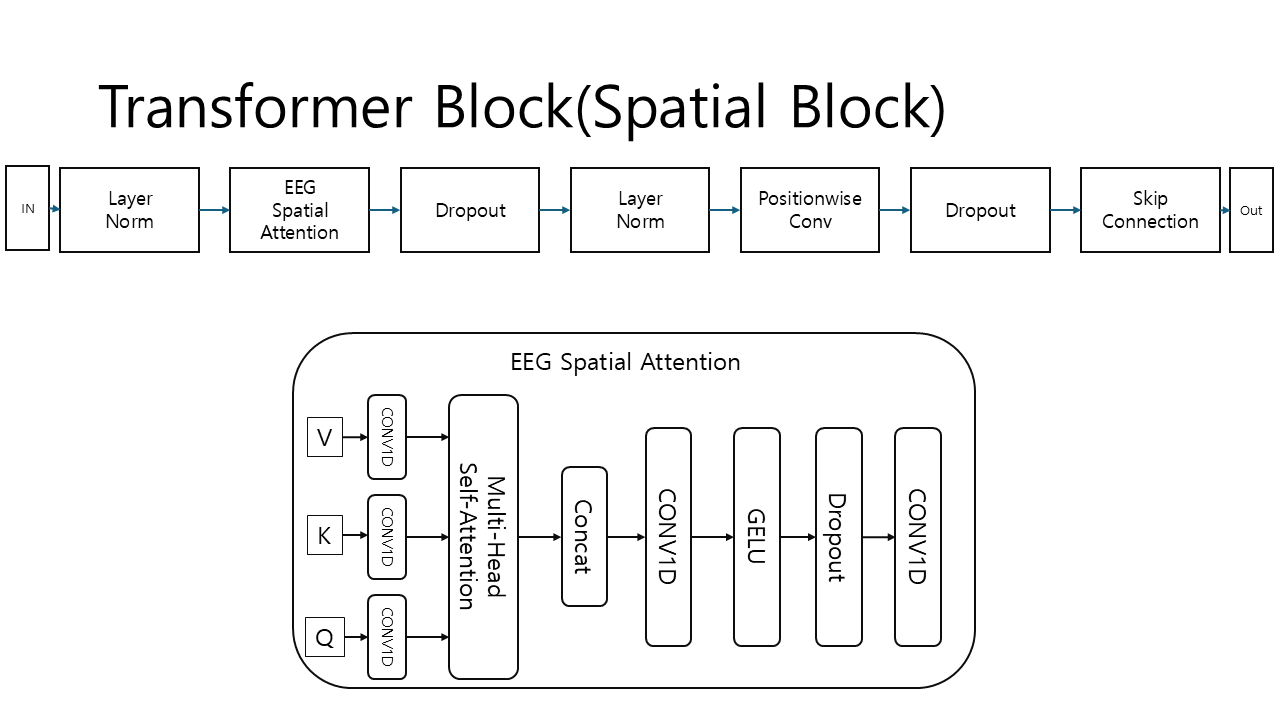}
\caption{Spatial Block with MHSA across EEG channels.}
\label{fig:spatial_block_sup}
\end{figure}

The Spatial Block (Figure~\ref{fig:spatial_block_sup}) applies multi-head self-attention (MHSA) across EEG channels, enabling the model to capture inter-regional correlations.

\section{Number of Trainable Parameters}
\label{sup:all_params}

\begin{table}[htbp]
  \centering
  \small
  \caption{Comparison of trainable parameter counts (in millions). 
  SYNAPSE achieves the lowest total parameters among all methods.}
  \label{tab:param_comparison}
  \begin{tabularx}{\linewidth}{@{} l *{3}{>{\centering\arraybackslash}X} @{}} 
    \toprule
    Model & EEG Module (M) ↓ & Conditioning (M) ↓ & Total (M) ↓ \\
    \midrule
    DreamDiffusion~\cite{dreamdiffusion} & 166.45 & 44.02 & 210.47 \\
    BrainVis~\cite{brainvis} & 171.10 & \textbf{24.82} & 194.92 \\
    GWIT~\cite{gwit} & \textbf{19.50} & 361.00 & 368.50 \\
    \textbf{Ours (SYNAPSE)} & 44.64 & 108.05 & \textbf{152.69} \\
    \bottomrule
  \end{tabularx}
\end{table}

\noindent
\textbf{EEG Module:} parameters for EEG feature extraction.  
For Ours and DreamDiffusion, this includes the encoder–decoder pair.  
For BrainVis and GWIT, this includes only the encoder.  
\textbf{Conditioning:} parameters for diffusion adaptation (e.g., mapping networks, ControlNet, or adapter layers).  

Compared with ControlNet-based frameworks such as GWIT~\cite{gwit}, 
SYNAPSE reduces total trainable parameters by 58\% (368M → 153M) while achieving a lower FID (46.9 vs.~80.5). 
This efficiency comes from (1) a compact CLIP-aligned EEG autoencoder that removes MAE pretraining redundancy, 
and (2) selective finetuning that updates only the adapter and key/value cross-attention layers.

\section{Retrieval Details}
\label{sup:retrieval_details}

We evaluate how well the EEG encoder aligns with the CLIP embedding space through two retrieval tasks: \textbf{image retrieval} and \textbf{label retrieval}. In both cases, cosine similarity is computed between EEG latent vectors and CLIP embeddings.

\paragraph{Local vs. Global Retrieval.}
The \textit{Local} setting measures retrieval within a validation batch (short-range context), 
while \textit{Global} evaluates across the entire dataset, testing generalization beyond the batch.

\begin{table}[h!]
  \centering
  \small
  \caption{Image Top-$K$ retrieval accuracy of the EEG encoder. 
  High Local scores indicate strong structural correspondence between EEG and visual features.}
  \label{tab:image_retrieval}
  \begin{tabularx}{\linewidth}{@{} l *{4}{>{\centering\arraybackslash}X} @{}} 
    \toprule
    Target & Subject & Top-1 ↑ & Top-3 ↑ & Top-5 ↑ \\
    \midrule
    Validation/Local   & 4         & 0.3527 & 0.6096 & 0.7762\\
    Validation/Global  & 4         & 0.0391 & 0.1423 & 0.2064\\
    \midrule
    Validation/Local   & General   & 0.4492 & 0.6867 & 0.7995\\
    Validation/Global  & General   & 0.0130 & 0.0554 & 0.1065\\
    Test/Local          & General   & 0.4261 & 0.6880 & 0.8028\\
    Test/Global         & General   & 0.0090 & 0.0482 & 0.1046\\
    \bottomrule
  \end{tabularx}
\end{table}

\paragraph{Label Top-$K$ Retrieval.}
In this task (Table~\ref{tab:label_retrieval}), we assess whether the retrieved images share the same \textit{class label} as the EEG’s ground truth.
This evaluates semantic-level consistency beyond exact image matching.

\begin{table}[h!]
  \centering
  \small
  \caption{Label Top-$K$ retrieval accuracy of the EEG encoder. 
  The Global setting demonstrates strong semantic generalization.}
  \label{tab:label_retrieval}
  \begin{tabularx}{\linewidth}{@{} l *{4}{>{\centering\arraybackslash}X} @{}} 
    \toprule
    Target & Subject & Top-1 $\uparrow$ & Top-3 $\uparrow$ & Top-5 $\uparrow$ \\
    \midrule
    Validation/Local   & 4         & 0.4330 & 0.6515 & 0.8112\\
    Validation/Global  & 4         & 0.4330 & 0.6515 & 0.8112\\
    \midrule
    Validation/Local   & General   & 0.4330 & 0.6515 & 0.8112\\
    Validation/Global  & General   & 0.5455 & 0.8450 & 0.9429\\
    Test/Local          & General   & 0.5122 & 0.7213 & 0.8320\\
    Test/Global         & General   & 0.6022 & 0.8742 & 0.9518\\
    \bottomrule
  \end{tabularx}
\end{table}

Under the Global setting, the encoder achieves up to 0.60 Top-1 and 0.95 Top-5 accuracy, confirming strong semantic correspondence to CLIP’s multimodal space.


\section{Full Autoencoder Results}
\label{sup:Autoencoder_full}

\begin{table}[htbp]
  \centering
  \small
  \caption{Detailed comparison of EEG autoencoder architectures. 
  Our CLIP-aligned model achieves the best reconstruction fidelity with substantially fewer parameters.}
  \label{tab:autoencoder_comparison_detailed}
  \resizebox{\textwidth}{!}{
  
  \begin{tabularx}{\textwidth}{X l X S[table-format=3.2] S[table-format=1.4] S[table-format=1.4]}
    \toprule
      Model & Subject & Latent Shape &
      \multicolumn{1}{c}{Params (M) $\downarrow$} &
      \multicolumn{1}{c}{MSE $\downarrow$} &
      \multicolumn{1}{c}{SDSC $\uparrow$} \\
     \midrule
    MAE~\cite{dreamdiffusion}     & 4       & 1x768    & 166.17 & 0.3040 & 0.3265 \\
    MAE~\cite{dreamdiffusion}     & General & 1x768    & 166.17 & 0.5570 & 0.2834 \\
    MAE~\cite{dreamdiffusion}     & 4       & 77x1024  & 166.45 & 0.2560 & 0.3788 \\
    MAE~\cite{dreamdiffusion}     & General & 77x1024  & 166.45 & 0.2403 & 0.7590 \\
    \midrule
    Ours (w/o align)              & 4       & 1x768    & {\bfseries 43.13} & 0.1720 & 0.6701 \\
    Ours (w/o align)              & General & 1x768    & {\bfseries 43.13} & 0.1204 & 0.7221 \\
    Ours (w/o align)              & 4       & 77x1024  & 44.63 & {\bfseries 0.0769} & {\bfseries 0.7955} \\
    Ours (w/o align)              & General & 77x1024  & 44.63 & 0.0810 & 0.7893 \\
    \midrule
    \bfseries Ours (CLIP-Aligned) & 4       & 77x1024  & 44.63 & 0.1085 & 0.7108 \\
    \bfseries Ours (CLIP-Aligned) & General & 77x1024  & 44.63 & 0.0882 & 0.7860 \\
    \bottomrule
  \end{tabularx}}
\end{table}

\noindent\textit{Interpretation.}
This table compares reconstruction fidelity (MSE↓, SDSC↑) and parameter counts across autoencoder variants.
The $77{\times}1024$ latent generally improves fidelity over $1{\times}768$.
Our model without alignment attains the best reconstruction, and the CLIP-aligned version shows only a small drop.

\noindent\textit{Takeaway.}
With \(\sim\)44.6M parameters versus 166.4M for MAE, our encoder is markedly more parameter-efficient while maintaining strong reconstruction, making it a suitable backbone for downstream alignment and generation.

\clearpage
\section{CFG Scale Test}
\label{sup:cfg_detail}

\begin{figure}[htbp]
\centering
\includegraphics[width=0.9\textwidth]{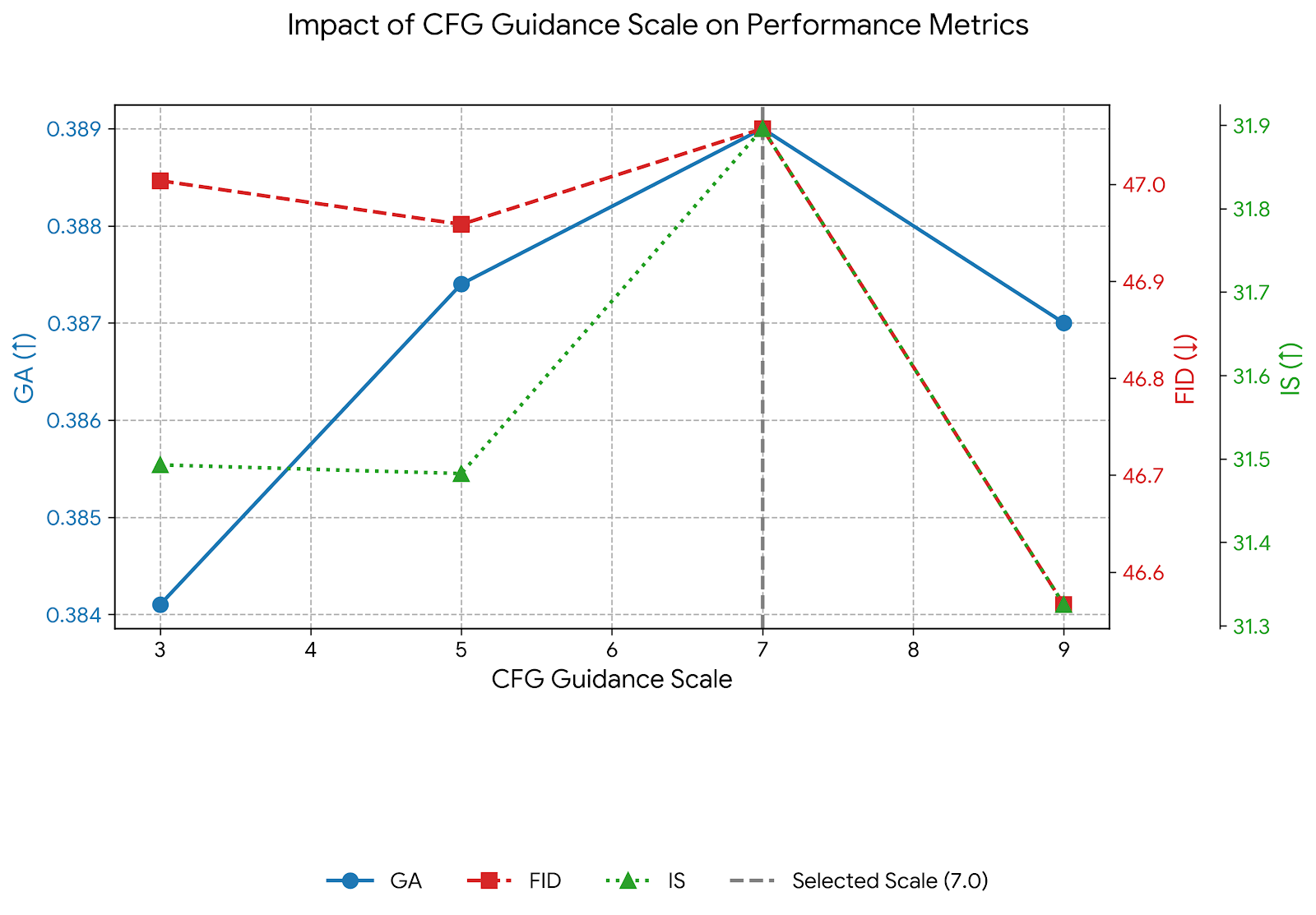}
\caption{Effect of CFG scale on FID and GA.}
\label{fig:cfg_detail}
\end{figure}

\begin{table}[htbp]
  \centering
  \small
  \caption{Comparison of CFG scales. 
  The best trade-off between fidelity and diversity occurs around 7–9, and we adopt 7.5 for all experiments.}
  \label{tab:cfg_results}
  \begin{tabular}{lccc}
    \toprule
    CFG Scale & FID $\downarrow$ & GA $\uparrow$ & IS $\uparrow$ \\
    \midrule
    3 & 47.00 & 0.3841 & 31.49 \\
    5 & 46.96 & 0.3874 & 31.48 \\
    7 & 47.06 & \textbf{0.3890} & \textbf{31.90} \\
    9 & \textbf{46.57} & 0.3870 & 31.33 \\
    \bottomrule
  \end{tabular}
\end{table}

\noindent\textit{Interpretation.}
We sweep classifier-free guidance (CFG) scales \(\{3,5,7,9\}\).
FID improves slightly toward 9, whereas GA/IS peak near 7, indicating a typical trade-off between fidelity and conditional adherence/diversity.

\noindent\textit{Protocol.}
Following common practice, we adopt CFG \(=7.5\) for all reported generations as a balanced operating point (details in \S\ref{sup:cfg_detail}).

\noindent\textit{Takeaway.}
Operating around 7–9 yields stable performance; \(7.5\) offers a robust default that balances FID and GA/IS without special tuning per class or subject.

\clearpage
\section{Cosine Analysis}

\noindent\textit{Description.}
Cosine similarity visualization across EEG-derived latent embeddings.
Each cell represents the pairwise similarity between mean-pooled latent vectors across all classes.

\noindent\textit{Observation.}
The diagonal dominance indicates strong intra-class coherence, while several off-diagonal regions (e.g., \textit{panda–elephant}, \textit{horse–monkey}) exhibit elevated similarity, revealing structured cross-class relations in the latent space.
These patterns are consistent with the visually plausible yet label-ambiguous generations observed in the main paper (Fig.~\ref{fig:bad_exam}).

\noindent\textit{Takeaway.}
The cosine map confirms that SYNAPSE preserves a semantically organized latent geometry:
even when categorical separability is imperfect, low- and mid-level visual features remain aligned across semantically related categories.
This structural consistency supports the claim that model errors arise from feature-level overlap rather than random noise.

\clearpage
\section{Case Examples}
\label{sup:more_examples}

\begin{figure}[htbp]
  \centering
  \includegraphics[width=\textwidth]{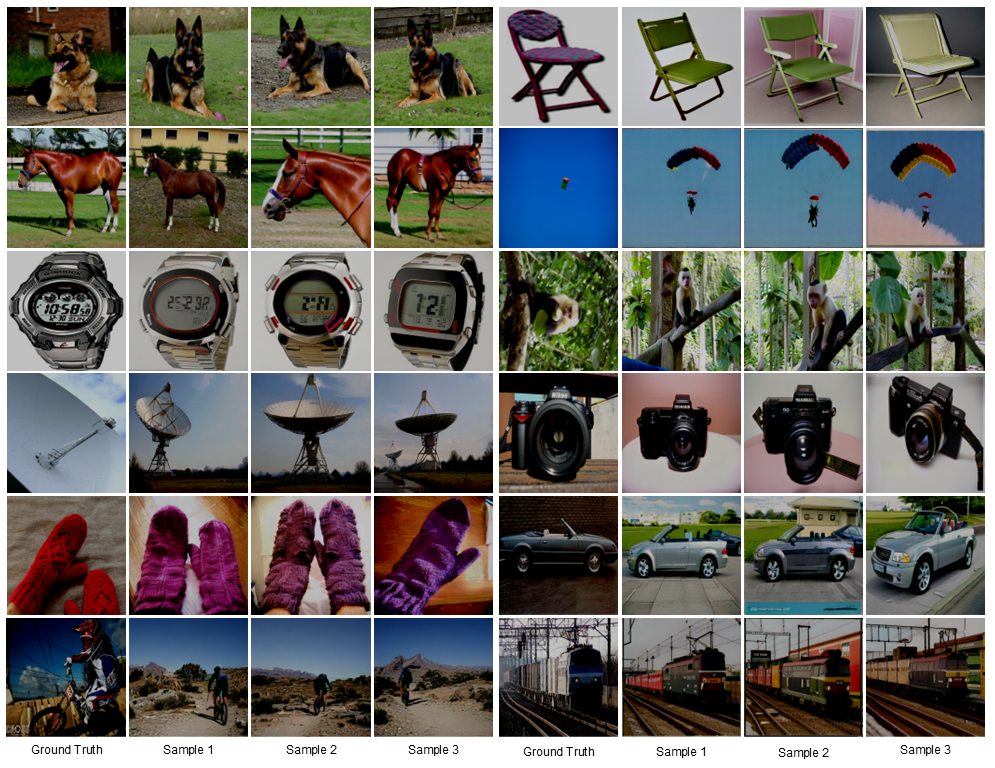}
  \caption{Representative good-case examples. SYNAPSE generates perceptually coherent and semantically aligned images.}
  \label{fig:good_exam}
\end{figure}

\noindent\textit{Description.}
Representative successful generations illustrating perceptual coherence and semantic consistency with the EEG-conditioned content (e.g., pose, context, background).

\noindent\textit{Observation.}
Across diverse categories, the samples capture fine-grained cues (shape, texture, scene context) rather than relying solely on class identity.

\noindent\textit{Takeaway.}
These qualitative examples complement the quantitative results by showing instance-level adherence to visual attributes aligned with the learned EEG latent.
Additional comparisons and mixed cases are provided in the supplementary qualitative section (\S\ref{sup:more_examples}).

\end{document}